\documentclass[a4paper]{jpconf}
\usepackage{graphicx}

\begin{document}
\rightline{WITS-MITP-002}
\vspace{1.8truecm}
\title{Two-loop renormalisation in UED models}

\author{Ammar Abdalgabar$^1$ and A. S. Cornell$^2$}

\address{National Institute for Theoretical Physics; School of Physics and Mandelstam Institute for Theoretical Physics,
University of the Witwatersrand, Wits 2050, South Africa}

\ead{$^1$ammar.abdalgabar@students.wits.ac.za, $^2$alan.cornell@wits.ac.za}

\begin{abstract}
The evolution equations of the gauge and Yukawa couplings  are derived for the two-loop renormalisation group equations in a five-dimensional SM compactified on a $S^1/Z_2$ to yield standard four space-time dimensions. Different possibilities can be discussed, however, we shall consider the limiting case in which all matter fields are localised on the brane. We will compare our two-loop results to the results found at one-loop level, and investigate the evolution of $\sin^2 \theta_W$ in this scenario also. 
\end{abstract}


\section{Introduction}

\par The renormalisation group evolution of gauge and Yukawa couplings is important in many contexts, and in some cases one needs to know this evolution quite precisely. Furthermore, in order to take full advantage of future experimental observations, it will be necessary to provide theoretical predictions for measured observables which are as accurate as possible, where one can hope that the values of these couplings will be deduced directly from experimental data.

\par Note that in order to obtain very precise predictions for the running of Yukawa couplings, one usually uses known masses of quarks and leptons \cite{Agashe:2014kda}, since it is the Yukawa interactions that give the fundamental fermions their masses after spontaneous symmetry breaking. Due to the observed hierarchy of the fermion masses, the corresponding values are usually defined at different scales. Therefore, one inevitably makes use of Renormalisation Group Equations (RGEs) to connect these scales. It should also be mentioned that contrary to leptons, quarks are not observed as free particles, so the pole mass which is usually associated with the physical mass of a particle, although being a gauge invariant and finite quantity, suffers from ambiguity \cite{Beneke:1994sw}. As a consequence, theoretical uncertainty in the determination of the top Yukawa coupling, for example, is reduced, thus requiring more precise determinations of the corresponding RGEs.

\par So by computing couplings to two-loop accuracy we are able to set limits, such as on the prediction of the masses of heavy fermions \cite{Machacek:1981ic}. But this is exasperated further when considering models with extra-dimensions, where we can have a range of heavy fermions and bosons, arising as Kaluza-Klein states, leading to runnings for our couplings now exhibiting a power law running, see Refs. \cite{Cornell:2010sz} for example. Furthermore, this observed power law running can mean contributions from higher-order loops could be very significant when compared to our earlier one-loop calculations \cite{Cornell:2010sz, Liu:2011gra}. Note that in this proceedings we shall restrict ourselves to a minimal Universal Extra-Dimensional (UED) model, where all matter fields are constrained to the brane, as a simpler testing ground for two-loop calculations. This shall be extended in an upcoming work to bulk cases also \cite{Upcoming}. However, two-loop precision remains desirable here because fermions can be heavy, and this means that their Yukawa couplings, and other couplings, can be large.

\par As such, the precise value of the corresponding coupling is required to test whether the Standard Model (SM), or UED models, correctly describe nature. The paper is organised as follows: In section 2 we shall recall the one- and two-loop SM beta functions. In section 3, the one- and two-loop UED beta functions shall be presented. Section 4 concludes with some preliminary results.


\section{The beta functions in the SM Model}

\par The one- and two-loop beta functions in the SM have been known for decades now in various mass limits, see Refs.~\cite{Fischler:1983sf} for example, and are presented here in the form we shall use for completeness.

\subsection{Gauge sector}

\par The one-loop beta functions for the gauge couplings in SM model are given by:
\begin{eqnarray}
(4 \pi)^2 g^{-3}_1\beta^{(1)}_{g_1}&=& \frac{41}{10}\,,\\
(4 \pi)^2 g^{-3}_2\beta^{(1)}_{g_2}&=& -\frac{19}{6}\,,\\
(4 \pi)^2 g^{-3}_3\beta^{(1)}_{g_3}&=& -7\,.
\end{eqnarray}

\noindent The two-loop contribution to the SM gauge couplings are given by:
\begin{eqnarray}
(4 \pi)^4 g^{-3}_1\beta^{(2)}_{g_1}&=& \frac{199}{50}g^2_1+\frac{27}{10}g^2_2+\frac{44}{5}g^2_3-\mathrm{Tr} \left(\frac{17}{5} Y_u^{\dagger} Y_u+ Y_d^{\dagger} Y_d+ 3Y_e^{\dagger} Y_e\right)\,,\\
(4 \pi)^4 g^{-3}_2\beta^{(2)}_{g_2}&=&\frac{9}{10}g^2_1+\frac{35}{6}g^2_2+12g^2_3-\mathrm{Tr} \left(3Y_u^{\dagger} Y_u+ 3Y_d^{\dagger} Y_d+ Y_e^{\dagger} Y_e\right)\,,\\
(4 \pi)^4 g^{-3}_3\beta^{(2)}_{g_3}&=& \frac{11}{10}g^2_1+\frac{9}{2}g^2_2-26g^2_3-4\mathrm{Tr} \left(Y_u^{\dagger} Y_u+ Y_d^{\dagger} Y_d\right)\,.
\end{eqnarray}

\subsection{Yukawa sector}

\par The one-loop beta function for the Yukawa couplings in SM model are given by:
\begin{eqnarray}
(4 \pi)^2 Y^{-1}_u\beta^{(1)}_{Y_u}&=& \mathrm{Tr} \left(3 Y_u^{\dagger} Y_u+3 Y_d^{\dagger} Y_d+ Y_e^{\dagger} Y_e\right)-\left(\frac{17}{20} g^2_1+\frac{9}{4} g^2_2+8 g^2_3\right) + \frac{3}{2}\left(Y_u^{\dagger} Y_u -Y_d^{\dagger}Y_d \right)\,,\\
(4 \pi)^2 Y^{-1}_d\beta^{(1)}_{Y_d}&=& \mathrm{Tr} \left(3 Y_u^{\dagger} Y_u+3 Y_d^{\dagger} Y_d+ Y_e^{\dagger} Y_e\right)-\left(\frac{1}{4} g^2_1+\frac{9}{4} g^2_2+8 g^2_3\right) + \frac{3}{2}\left(Y_d^{\dagger} Y_d -Y_u^{\dagger}Y_u\right)\,,\\
(4 \pi)^2 Y^{-1}_e\beta^{(1)}_{Y_e}&=& \mathrm{Tr} \left(3 Y_u^{\dagger} Y_u+3 Y_d^{\dagger} Y_d+ Y_e^{\dagger} Y_e\right)-\left(\frac{9}{4} g^2_1+\frac{9}{4} g^2_2\right) + \frac{3}{2}Y_e^{\dagger} Y_e\,.
\end{eqnarray}

\noindent The two-loop contribution to the SM Yukawa couplings are given by:
\begin{eqnarray}
(4 \pi)^4 Y^{-1}_u\beta^{(2)}_{Y_u}&=& 11(Y_d^{\dagger} Y_d)^2-5(Y_d^{\dagger} Y_d)(Y_u^{\dagger} Y_u)+\left(5Y_d^{\dagger} Y_d-9Y_u^{\dagger} Y_u \right)\mathrm{Tr} \left(3 Y_u^{\dagger} Y_u+3 Y_d^{\dagger} Y_d+ Y_e^{\dagger} Y_e\right)\nonumber\\
&&6(Y_u^{\dagger} Y_u)^2-9\mathrm{Tr} \left(3 (Y_u^{\dagger} Y_u)^2+3( Y_d^{\dagger} Y_d)^2-\frac{2}{3}(Y_d^{\dagger} Y_d)(Y_u^{\dagger} Y_u)+ (Y_e^{\dagger} Y_e)^2\right)\nonumber\\
&&+Y_u^{\dagger} Y_u\left(\frac{223}{40} g^2_1+\frac{135}{8} g^2_2+32 g^2_3\right)-Y_d^{\dagger} Y_d\left(\frac{43}{40} g^2_1-\frac{9}{8} g^2_2+32 g^2_3\right)\nonumber\\
&&+\frac{1187}{600}g^4_1-\frac{23}{4}g^4_2-108g^4_3-\frac{9}{20}g^2_1g^2_2-\frac{19}{15}g^2_1g^2_3+9g^2_2g^2_3+ \frac{15}{4}\mathrm{Tr}(Y_e^{\dagger} Y_e)\left(g^2_1+g^2_2\right)\nonumber\\
&& + 5\mathrm{Tr}(Y_u^{\dagger} Y_u)\left(\frac{17}{20} g^2_1+\frac{9}{4} g^2_2+8 g^2_3\right)+ 5\mathrm{Tr}(Y_d^{\dagger} Y_d)\left(\frac{1}{4} g^2_1+\frac{9}{4} g^2_2+8 g^2_3\right)\,,
\end{eqnarray}
\begin{eqnarray}
(4 \pi)^4 Y^{-1}_d\beta^{(2)}_{Y_d}&=& 11(Y_u^{\dagger} Y_u)^2-5(Y_d^{\dagger} Y_d)(Y_u^{\dagger} Y_u)+\left(5Y_u^{\dagger} Y_u-9Y_d^{\dagger} Y_d \right)\mathrm{Tr} \left(3 Y_u^{\dagger} Y_u+3 Y_d^{\dagger} Y_d+ Y_e^{\dagger} Y_e\right)\nonumber\\
&&6(Y_d^{\dagger} Y_d)^2-9\mathrm{Tr} \left(3 (Y_u^{\dagger} Y_u)^2+3( Y_d^{\dagger} Y_d)^2-\frac{2}{3}(Y_d^{\dagger} Y_d)(Y_u^{\dagger} Y_u)+ (Y_e^{\dagger} Y_e)^2\right)\nonumber\\
&&+Y_d^{\dagger} Y_d\left(\frac{187}{40} g^2_1+\frac{135}{8} g^2_2+32 g^2_3\right)-Y_u^{\dagger} Y_u\left(\frac{79}{40} g^2_1-\frac{9}{8} g^2_2+32 g^2_3\right)\nonumber\\
&&-\frac{127}{600}g^4_1-\frac{23}{4}g^4_2-108g^4_3-\frac{27}{20}g^2_1g^2_2+\frac{13}{15}g^2_1g^2_3+9g^2_2g^2_3+ \frac{15}{4}\mathrm{Tr}(Y_e^{\dagger} Y_e)\left(g^2_1+g^2_2\right)\nonumber\\
&& + 5\mathrm{Tr}(Y_u^{\dagger} Y_u)\left(\frac{17}{20} g^2_1+\frac{9}{4} g^2_2+8 g^2_3\right)+ 5\mathrm{Tr}(Y_d^{\dagger} Y_d)\left(\frac{1}{4} g^2_1+\frac{9}{4} g^2_2+8 g^2_3\right)\,,
\end{eqnarray}
\begin{eqnarray}
(4 \pi)^4 Y^{-1}_e\beta^{(2)}_{Y_e}&=&6(Y_e^{\dagger} Y_e)^2-9\mathrm{Tr} \left(3 (Y_u^{\dagger} Y_u)^2+3( Y_d^{\dagger} Y_d)^2-\frac{2}{3}(Y_d^{\dagger} Y_d)(Y_u^{\dagger} Y_u)+ (Y_e^{\dagger} Y_e)^2\right)\nonumber\\
&&-9Y_e^{\dagger}Y_e \mathrm{Tr} \left(3 Y_u^{\dagger} Y_u+3 Y_d^{\dagger} Y_d+ Y_e^{\dagger} Y_e\right)+Y_e^{\dagger} Y_e\left(\frac{387}{40} g^2_1+\frac{135}{8} g^2_2\right)\nonumber\\
&& + 5\mathrm{Tr}(Y_u^{\dagger} Y_u)\left(\frac{17}{20} g^2_1+\frac{9}{4} g^2_2+8 g^2_3\right)+ 5\mathrm{Tr}(Y_d^{\dagger} Y_d)\left(\frac{1}{4} g^2_1+\frac{9}{4} g^2_2+8 g^2_3\right)\nonumber\\
&&+\frac{15}{4}\mathrm{Tr}(Y_e^{\dagger} Y_e)\left(g^2_1+g^2_2\right)+\frac{1371}{200}g^4_1-\frac{23}{4}g^4_2+\frac{27}{20}g^2_1g^2_2\,.
\end{eqnarray}


\section{The beta functions in a brane-localised UED model}

\par In this scenario the SM $SU(3)_c\times SU(2)_L\times U(1)_Y$ gauge fields and the Higgs ($H$) propagate into the fifth dimension, $y$. As a consequence these fields will have Kaluza-Klein modes which contribute to the RGEs at energies $E> 1/R$. Different possibilities for the matter fields can be studied, however we shall study the limiting case with SM matter fields restricted to the $y=0$ brane. As such there will be no additional Kaluza-Klein contributions of these matter fields to the RGEs.

\subsection{Gauge sector}

\par The one-loop beta functions for the gauge couplings in this five dimensional model are given by:
\begin{eqnarray}
(4 \pi)^2 g^{-3}_1\beta^{(1)}_{g_1}&=& (S(t)-1)\Big(\frac{1}{10}+\frac{8}{3}\eta\Big)\,,\\
(4 \pi)^2 g^{-3}_2\beta^{(1)}_{g_2}&=& (S(t)-1)\Big(-\frac{41}{6}+\frac{8}{3}\eta\Big)\,,\\
(4 \pi)^2 g^{-3}_3\beta^{(1)}_{g_3}&=& (S(t)-1)\Big(-\frac{21}{2}+\frac{8}{3}\eta\Big)\,.
\end{eqnarray}
where $\eta$ is the number of fermion generations.

\par The two-loop contribution to the gauge couplings in this five dimensional model are given by:
\begin{eqnarray}
(4 \pi)^4 g^{-3}_1\beta^{(2)}_{g_1}&=&(S(t)^2-1)\left(\frac{199}{50}g^2_1+\frac{27}{10}g^2_2+\frac{44}{5}g^2_3\right)\,,\\
(4 \pi)^4 g^{-3}_2\beta^{(2)}_{g_2}&=&(S(t)^2-1)\left(\frac{9}{10}g^2_1+\frac{35}{6}g^2_2+12g^2_3\right)\,,\\
(4 \pi)^4 g^{-3}_3\beta^{(2)}_{g_3}&=&(S(t)^2-1)\left(\frac{11}{10}g^2_1+\frac{9}{2}g^2_2-26g^2_3\right)\,.
\end{eqnarray}

\subsection{Yukawa sector}

\par The one-loop beta functions for the Yukawa couplings in this five dimensional model are given by:
\begin{eqnarray}
(4 \pi)^2 Y^{-1}_u\beta^{(1)}_{Y_u}&=& 2(S(t)-1)\left(-\left(\frac{17}{20} g^2_1+\frac{9}{4} g^2_2+8g^2_3\right)+\frac{3}{2}\left(Y_u^{\dagger} Y_u -Y_d^{\dagger}Y_d \right)\right)\,,\\
(4 \pi)^2 Y^{-1}_d\beta^{(1)}_{Y_d}&=& 2(S(t)-1)\left(-\left(\frac{1}{4} g^2_1+\frac{9}{4} g^2_2+8g^2_3\right)+ \frac{3}{2}\left(Y_d^{\dagger} Y_d -Y_u^{\dagger}Y_u\right)\right)\,,\\
(4 \pi)^2 Y^{-1}_e\beta^{(1)}_{Y_e}&=& 2(S(t)-1)\left(-\left(\frac{9}{4} g^2_1+\frac{9}{4} g^2_2\right) + \frac{3}{2}Y_e^{\dagger} Y_e\right)\,.
\end{eqnarray}

\noindent The two-loop contribution to the Yukawa couplings in this five dimensional are given by:
\begin{eqnarray}
(4 \pi)^4 Y^{-1}_u\beta^{(2)}_{Y_u}&=& 2(S(t)^2-1)\left(6(Y_u^{\dagger} Y_u)^2-5(Y_d^{\dagger} Y_d)(Y_u^{\dagger} Y_u)+11(Y_d^{\dagger} Y_d)^2\right.\nonumber\\
&&\left.+Y_u^{\dagger} Y_u\left(\frac{223}{40} g^2_1+\frac{135}{8} g^2_2+32 g^2_3\right)-Y_d^{\dagger} Y_d\left(\frac{43}{40} g^2_1-\frac{9}{8} g^2_2+32 g^2_3\right)\right.\nonumber\\
&&\left.+\frac{1187}{600}g^4_1-\frac{23}{4}g^4_2-108g^4_3-\frac{9}{20}g^2_1g^2_2-\frac{19}{15}g^2_1g^2_3+9g^2_2g^2_3\right)\,,\\
(4 \pi)^4 Y^{-1}_d\beta^{(2)}_{Y_d}&=& 2(S(t)^2-1)\left(6(Y_d^{\dagger} Y_d)^2-5(Y_d^{\dagger} Y_d)(Y_u^{\dagger} Y_u)+11(Y_u^{\dagger} Y_u)^2\right.\nonumber\\
&&\left.+Y_d^{\dagger} Y_d\left(\frac{187}{40} g^2_1+\frac{135}{8} g^2_2+32 g^2_3\right)-Y_u^{\dagger} Y_u\left(\frac{79}{40} g^2_1-\frac{9}{8} g^2_2+32 g^2_3\right)\right.\nonumber\\
&&\left.-\frac{127}{600}g^4_1-\frac{23}{4}g^4_2-108g^4_3-\frac{27}{20}g^2_1g^2_2+\frac{13}{15}g^2_1g^2_3+9g^2_2g^2_3\right)\,,\\
(4 \pi)^4 Y^{-1}_e\beta^{(2)}_{Y_e}&=& 2(S(t)^2-1)\left(6(Y_e^{\dagger} Y_e)^2+Y_e^{\dagger} Y_e\left(\frac{387}{40} g^2_1+\frac{135}{8} g^2_2\right)\right.\nonumber\\
&&\left.+\frac{1371}{200}g^4_1-\frac{23}{4}g^4_2+\frac{27}{20}g^2_1g^2_2\right)\,.
\end{eqnarray}
That is, when the energy scale parameter $t=\ln(E/M_Z)>\ln(1/R M_Z)$ or $E>1/R$, these equations shall be used. When the energy $0<t<\ln(1/R M_Z)$ (that is $M_Z<E<1/R$) the evolution of the gauge and Yukawa couplings at low energy are given by the usual four dimensional SM expressions, as was presented in section 2.


\section{Result and Discussion}

\begin{figure}[th]
\begin{center}
\includegraphics[width=10cm,angle=0]{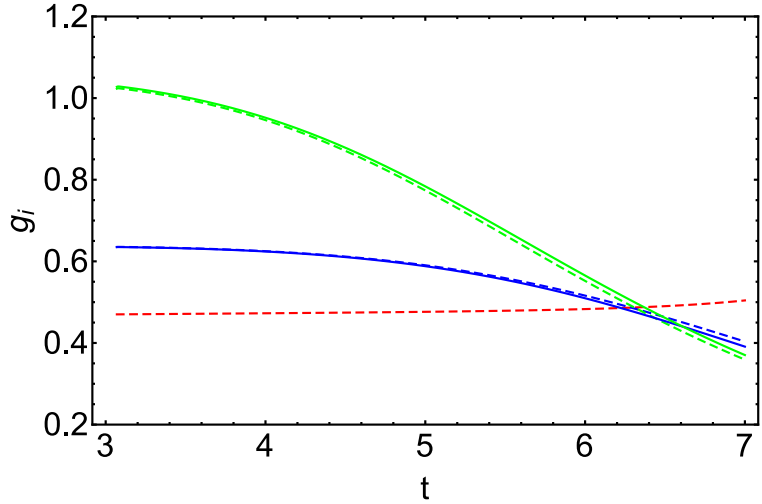} 
\end{center}
\caption{\it (Colour online) Gauge couplings {$g_1$} (red), {$g_2$} (blue), {$g_3$} (green) with all matter fields in the brane; for the compactification scales $R^{-1} = 2$ TeV as a function of the scale parameter {$t$}. Solid lines are the one-loop level runnings, dashed lines are two-loop level runnings.}
\label{fig:gauge}
\end{figure}
\begin{figure}[h!]
\begin{center}
\includegraphics[width=10cm,angle=0]{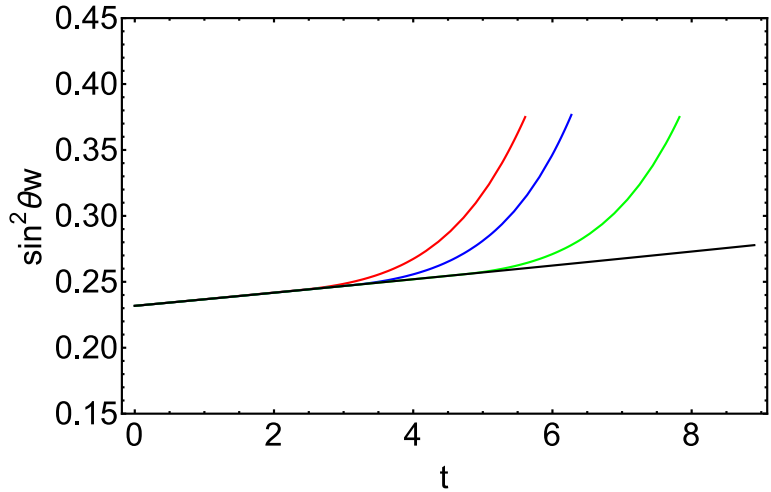} 
\end{center}
\caption{\it $sin^2 \theta_W$ with all matter fields in the brane; for three different values of the compactification scales $R^{-1} = 1$ TeV (red), 2 TeV (blue), 10 TeV (green) as a function of the scale parameter {$t$}.}
\label{fig:Weinberg}
\end{figure}
\begin{figure}[h!]
\begin{center}
\includegraphics[width=10cm,angle=0]{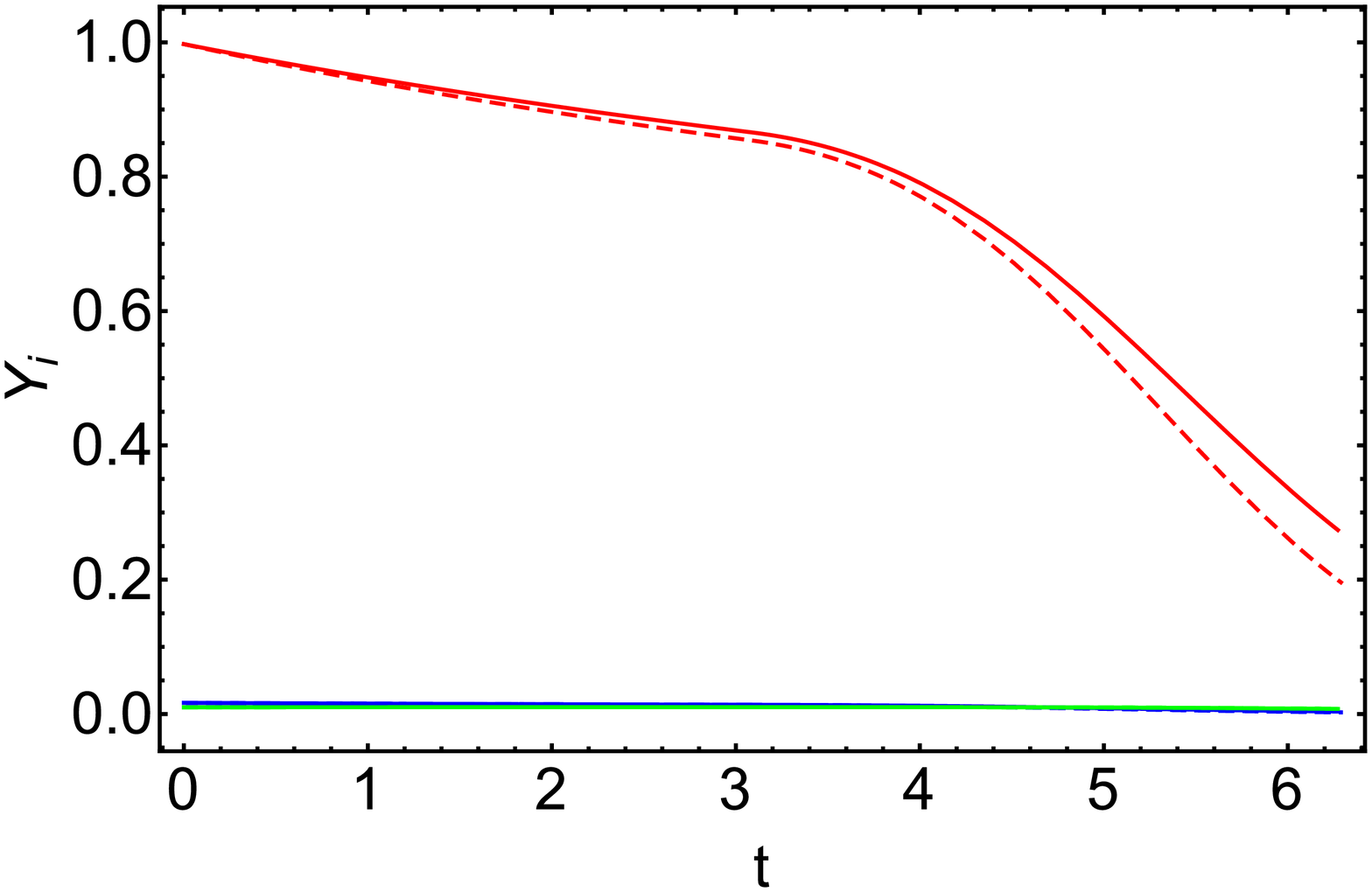} 
\end{center}
\caption{\it (Colour online) Yukawa couplings {$Y_e$} (green), {$Y_d$} (blue), {$Y_u$} (red) with: all matter fields in the brane; for the compactification scales $R^{-1} = 2$ TeV as a function of the scale parameter {$t$}. As in Fig.~\ref{fig:gauge}, solid lines are the one-loop level runnings, dashed lines are two-loop level runnings.}
\label{fig:Yukawas}
\end{figure}

\par As expected the extra-dimension brings down the unification scale to a lower scale, see Fig.~\ref{fig:gauge}. At one-loop level the gauge and Yukawa coupling RGEs are disentangled, however, at the two-loop level the RGEs for the gauge and Yukawa couplings are entangled. As such, only a few percent change in the evolution of gauge couplings, due to the appearance of Yukawa couplings over the one-loop running, is observed. This also may change the running of $g_3$, due to  the large size of $Y_t$. 

\par In Fig.~\ref{fig:Weinberg} we present the evolution of $\sin^2 \theta_W$ in this brane constrained UED model. Once the new contributions from the extra-dimensions arise, the behaviour is changed until we reach the cut-off scale. One can see for $R^{-1} = 1$ TeV $\sin^2 \theta_W$ can rise to $\sim 0.4$. This result may be useful, at least from a model building perspective, where there is no discernible difference from the one- and two-loop trajectories in this plot.

\par As such, for models with an extra-dimension the one-loop running of Yukawa couplings is clearly insufficient, since higher order corrections can be just as important at scales a few times above $1/R$. Although this type of large correction, which we require to claim unification, see Fig.~\ref{fig:Yukawas}, needs to be tested further to ensure that $Y_t$ remains perturbative up to the unification scale. This shall be more fully explored in an upcoming work, where the case of bulk propagating matter fields shall also be investigated \cite{Upcoming}.


\section*{Acknowledgments}

We would like to thank our collaborator Aldo Deandrea for the helpful discussions. This work is supported by the National Research Foundation (South Africa).


\end{document}